\documentclass[12pt]{iopart}
\usepackage{iopams}  
\usepackage{epsfig}

\usepackage{bbm}
\usepackage{color}
\usepackage{ulem}

\usepackage{enumitem}
\usepackage{graphicx}
\usepackage{subfigure}

\newcommand{\be}{\begin{equation}}
\newcommand{\ee}{\end{equation}}
\newcommand{\bea}{\begin{eqnarray}}
\newcommand{\eea}{\end{eqnarray}}
\newcommand{\nn}{\nonumber}

\def\({\left(}
\def\){\right)}

\newcommand{\half}{\frac{1}{2}}

\begin{document}

\title{Generalizing Galileons}

\author{Mark Trodden and Kurt Hinterbichler}

\address{Center for Particle Cosmology, Department of Physics and Astronomy, University of Pennsylvania,
Philadelphia, Pennsylvania 19104, USA}
\ead{trodden@physics.upenn.edu}
\ead{kurthi@physics.upenn.edu}
\begin{abstract}
The Galileons are a set of terms within four-dimensional effective field theories, obeying symmetries that can be derived from the
dynamics of a $3+1$-dimensional flat brane embedded in a $5$-dimensional Minkowski Bulk. These theories have some intriguing properties, including freedom from ghosts and a non-renormalization theorem that hints at possible applications in both particle physics and cosmology. In this brief review article, we will summarize our attempts over the last year to extend the Galileon idea in two important ways. We will discuss the effective field theory construction arising from co-dimension greater than one flat branes embedded in a flat background - the multiGalileons - and we will then describe symmetric covariant versions of the Galileons, more suitable for general cosmological applications. While all these Galileons can be thought of as interesting four-dimensional field theories in their own rights, the work described here may also make it easier to embed them into string theory, with its multiple extra dimensions and more general gravitational backgrounds.
\end{abstract}

\maketitle

\section{Introduction}
The Dvali-Gabadadze-Porrati (DGP) model~\cite{Dvali:2000hr}, has provided an interesting playground within which cosmologists and particle physicists can explore new approaches to cosmology, to modifying gravity, and to constructing new four-dimensional effective field theories. The DGP model consists of a $3+1$-dimensional brane, embedded in a $5$d bulk, with the unusual, and deceptively simple action
\be
S=\frac{M_5^3}{2}\int d^5X\, \sqrt{-G}\,\  R[G] + \frac{M_4^2}{2}\int d^4x\, \sqrt{-g}\, \ R[g] \ .
\label{DGPaction}
\ee
This model yields a rich and dramatic phenomenology. Gravity is modified on large distances, and this modification allows a branch of $4$-dimensional cosmological solutions which self-accelerate at late times. As a result, the DGP model has been the subject of a great deal of scrutiny, resulting in a number of important constraints; most importantly the fact that the accelerating branch inevitably gives rise to a ghost.

Putting the desire for self acceleration aside for a moment, the normal branch is healthy, and it is possible to derive a 4d effective action for the DGP model by integrating out the bulk.  It has been claimed~\cite{Luty:2003vm,Nicolis:2004qq} (see also \cite{Gabadadze:2006tf}) that a decoupling limit for DGP then exists, in which the 4d effective action reduces to a theory of a single scalar $\pi$, representing the position of the brane in the extra dimension, with a cubic self-interaction term $\sim (\partial\pi)^2\Box \pi$.  This term has the properties that its field equations are second order (despite the fact that the Lagrangian is higher order).   It is also invariant (up to a total derivative) under the following {\it Galilean} transformations
\be 
\label{Galileoninvarianceold}
\pi (x) \rightarrow \pi(x) + c + b_\mu x^\mu \ ,
\ee
where $c$ and $b_{\mu}$ are constants.  
This symmetry is inherited from a combination of five dimensional Poincar\'e invariance and brane reparametrization invariance in a small field limit, and has been called the {\it Galilean} symmetry (due to its similarities to the Galilean boost of a non-relativistic point particle).  

It is possible to abstract this structure, to consider a four dimensional field theory with this same symmetry, and with the associated scalar becoming the {\it Galileon}~\cite{Nicolis:2008in}.  Interestingly, there are a finite number of terms, the {\it Galileon terms}, that have fewer numbers of derivatives per field than the infinity of competing terms with the same symmetries.  The Galileon terms have the surprising property that, despite the presence of higher derivatives in the actions, the equations of motion are second order, so that no extra degrees of freedom are propagated around any background.  Much is now known about these, including a non-renormalization theorem~\cite{Luty:2003vm,Hinterbichler:2010xn,Burrage:2010cu}; cosmological applications~\cite{Burrage:2010cu,Agarwal:2011mg,Creminelli:2010ba,Creminelli:2010qf,DeFelice:2010as,Deffayet:2010qz,Kobayashi:2011pc,Mota:2010bs,Wyman:2011mp}; covariantizations~\cite{Deffayet:2009mn,Deffayet:2009wt,Deffayet:2011gz,Goon:2011qf,Goon:2011uw,Burrage:2011bt}; and extensions to p-forms~\cite{Deffayet:2010zh} and supersymmetry~\cite{Khoury:2011da}.  

Abstractions aside, within the context of brane constructions theories of this sort are generic since they share, in a certain limit, the symmetries of the Dirac-Born-Infeld (DBI) action.  The DBI action encodes the lowest order dynamics of a brane embedded in higher dimensions. The Galileon terms can be thought of as a subset of the higher order terms expected to be present in any effective field theory of the brane, and which will be suppressed by powers of some cutoff scale.  This subset is special because its members contain fewer derivatives per field than competing terms with the same symmetries, and because they yield second order equations.

Crucially, due to the fact that the Galileons contain fewer derivatives that other terms, there can exist regimes in which only the finite number of Galileon terms are important, and the infinity of other possible terms within the effective field theory are not (see section II of \cite{Hinterbichler:2010xn}, as well as \cite{Nicolis:2004qq,Endlich:2010zj}, for more on this and for examples of such regimes.)  This fact, coupled with a non-renormalization theorem for Galileons, holds out the hope of computing non-linear facts about the world which are exact quantum mechanically.  It should be remembered that even if our universe is not a brane world, the same conclusions follow if one postulates the existence of symmetries of the same form as those of a brane world.

In this article we will review work done over the last year in which we have extended the Galileon idea in two important directions. We will discuss the structure of mutli-Galileon theories~\cite{Hinterbichler:2010xn,Padilla:2010de,Padilla:2010ir,Padilla:2010tj,Zhou:2010di}, and describe how they, and the symmetries they exhibit, can arise from embedding branes in multiple co-dimensions. We will then discuss how to covariantize the Galileon model in a way which preserves the symmetries. This leads to entirely new four-dimensional effective field theories with the same numbers of symmetries as the Galileon theories, but with different structures, allowing for fields propagating on curved space, and allowing for potentials with implications for cosmology and for particle physics.

%%%%%%%%%%%%%%%%%%%%%%%%%%%%%

\section{The Galileon: A Review}

As mentioned in the introduction, the decoupling limit of DGP is a $4$-dimensional effective theory of gravity coupled to a single scalar field $\pi$, the bending mode of the brane in the fifth dimension. Nicolis, Rattazzi and Trincherini~\cite{Nicolis:2008in} generalized this structure to all possible terms for a single scalar with the same properties.  They found that there exists a single Galileon term at each order in $\pi$, where ``order" refers to the number of copies of $\pi$ that appear in the term.  For $n\geq 1$, the $(n+1)$-th order Galileon Lagrangian is
\be
\label{Galileon2} 
{\cal L}_{n+1}=n\eta^{\mu_1\nu_1\mu_2\nu_2\cdots\mu_n\nu_n}\left( \partial_{\mu_1}\pi\partial_{\nu_1}\pi\partial_{\mu_2}\partial_{\nu_2}\pi\cdots\partial_{\mu_n}\partial_{\nu_n}\pi\right) \ ,
\ee 
where 
\be
\label{tensor} 
\eta^{\mu_1\nu_1\mu_2\nu_2\cdots\mu_n\nu_n}\equiv{1\over n!}\sum_p\left(-1\right)^{p}\eta^{\mu_1p(\nu_1)}\eta^{\mu_2p(\nu_2)}\cdots\eta^{\mu_np(\nu_n)} \ .
\ee 
The sum in~(\ref{tensor}) is over all permutations of the $\nu$ indices, with $(-1)^p$ the sign of the permutation.  The tensor~(\ref{tensor}) is anti-symmetric in the $\mu$ indices, anti-symmetric the $\nu$ indices, and symmetric under interchange of any $\mu,\nu$ pair with any other.  These Lagrangians are unique up to total derivatives and overall constants.   Because of the anti-symmetry requirement on $\eta$, only the first $n$ of these Galileons are non-trivial in $n$-dimensions.  In addition, the tadpole term, $\pi$, is Galilean invariant, and we therefore include it as the first-order Galileon.  

Thus, at the first few orders, we have 
\bea
 {\cal L}_1&=&\pi, \\ \nn
 {\cal L}_2&=&[\pi^2], \\ \nn
{\cal L}_3&=&[\pi^2][\Pi]-[\pi^3], \\ \nn
{\cal L}_4&=&\half[\pi^2][\Pi]^2-[\pi^3][\Pi]+[\pi^4]-\half[\pi^2][\Pi^2], \\ \nn
{\cal L}_5&=&{1\over 6}[\pi^2][\Pi]^3-{1\over 2}[\pi^3][\Pi]^2+[\pi^4][\Pi]-[\pi^5]+{1\over 3}[\pi^2][\Pi^3] \nonumber \\
&& -{1\over 2}[\pi^2][\Pi][\Pi^2]+{1\over 2}[\pi^3][\Pi^2]
\ .
\eea
We have used the notation $\Pi$ for the matrix of partials $\Pi_{\mu\nu}\equiv\partial_{\mu}\partial_\nu\pi$, and $[\Pi^n]\equiv Tr(\Pi^n)$, e.g. $[\Pi]=\square\pi$, $[\Pi^2]=\partial_\mu\partial_\nu\pi\partial^\mu\partial^\nu\pi$, and $[\pi^n]\equiv \partial\pi\cdot\Pi^{n-2}\cdot\partial\pi$, i.e. $[\pi^2]=\partial_\mu\pi\partial^\mu\pi$, $[\pi^3]=\partial_\mu\pi\partial^\mu\partial^\nu\pi\partial_\nu\pi$.  The above terms are the only ones which are non-vanishing in four dimensions.  The second is the standard kinetic term for a scalar, while the third is the DGP $\pi$-Lagrangian (up to a total derivative).

The equations of motion derived from (\ref{Galileon2}) are
\be 
\label{Galileoneom} 
{\cal E}_{n+1} \equiv -n(n+1)\eta^{\mu_1\nu_1\mu_2\nu_2\cdots\mu_n\nu_n}\left( \partial_{\mu_1}\partial_{\nu_1}\pi\partial_{\mu_2}\partial_{\nu_2}\pi\cdots\partial_{\mu_n}\partial_{\nu_n}\pi\right)=0 \ ,
\ee
and are second order, so the scalar does not propagate extra ghostly degrees of freedom. The equations of motion are 
\bea{\cal E}_1&=&1, \\
 {\cal E}_2&=&-2[\Pi], \\
{\cal E}_3&=& -3\left([\Pi]^2-[\Pi^2]\right),  \\
{\cal E}_4&=& -2\left([\Pi]^3+2[\Pi^3]-3[\Pi][\Pi^2]\right),  \\
{\cal E}_5&=& -{5\over 6}\left([\Pi]^4-6[\Pi^4]+8[\Pi][\Pi^3]-6[\Pi]^2[\Pi^2]+3[\Pi^2]^2\right) \ ,
\eea
and are purely second order in derivatives.
By adding a total derivative, and by using the following identity for the $\eta$ symbol in ${\cal L}_{n+1}$
\bea
\eta^{\mu_1\nu_1\ldots\mu_n\nu_n}={1\over n}\left(\eta^{\mu_1\nu_1}\eta^{\mu_2 \nu_2\ldots\mu_n \nu_n} \right.&-&\eta^{\mu_1\nu_2}\eta^{\mu_2 \nu_1\mu_2\nu_3\ldots\mu_n\nu_n}+\cdots \nonumber \\
&+& \left. (-1)^n\eta^{\mu_1\nu_n}\eta^{\mu_2\nu_1\ldots\mu_n\nu_{n-1}}\right) \ ,
\eea
the Galileon Lagrangians can be brought into a (sometimes more useful) different form, which illustrates that the $(n+1)$-th order Lagrangian is just $(\partial\pi)^2$ times the $n$-th order equations of motion,
\bea
{\cal L}_{n+1}= &-&{n+1\over 2n(n-1)}(\partial\pi)^2{\cal E}_{n} \nonumber \\
&-&{n-1\over 2}\partial_{\mu_1}\left[(\partial\pi)^2\eta^{\mu_1\nu_1\cdots\mu_{n-1}\nu_{n-1}}\partial_{\nu_1}\pi\partial_{\mu_2}\partial_{\nu_2}\pi\cdots\partial_{\mu_{n-1}}\partial_{\nu_{n-1}}\pi\right] \ .
\label{simplifiedGalileon}
\eea
From the simplified form~(\ref{simplifiedGalileon}) we can see that ${\cal L}_3$, for example, takes the usual Galileon form $(\partial\pi)^2\Box\pi$.  This also connects to the Euler-heirarchy constructions of \cite{Fairlie:1991qe,Fairlie:1992nb,Fairlie:1992yy,Fairlie:2011md}.

These Galileon actions can be generalized to the multi-field case, where there is a multiplet $\pi^I$ of fields~\cite{Deffayet:2010zh,Padilla:2010de,Padilla:2010ir,Hinterbichler:2010xn,Zhou:2010di}.  The action in this case can be written
\be 
\label{generaltermspre} 
{\cal L}_{n+1}= S_{I_1I_2\cdots I_{n+1}}\eta^{\mu_1\nu_1\mu_2\nu_2\cdots\mu_n\nu_n}\left(\pi^{I_{n+1}} \partial_{\mu_1}\partial_{\nu_1}\pi^{I_1}\partial_{\mu_2}\partial_{\nu_2}\pi^{I_2}\cdots\partial_{\mu_n}\partial_{\nu_n}\pi^{I_{n}}\right) \ ,
\ee
with $S_{I_1I_2\cdots I_{n+1}}$ a symmetric constant tensor.  This is invariant under under individual Galilean transformations for each field, $\pi^I\rightarrow \pi^I+c^I+b^I_\mu x^\mu$, and the equations of motion are second order,
\be 
{\cal E}_{I}\equiv (n+1)S_{II_1I_2\cdots I_{n}}\eta^{\mu_1\nu_1\mu_2\nu_2\cdots\mu_n\nu_n}\left(\partial_{\mu_1}\partial_{\nu_1}\pi^{I_1}\partial_{\mu_2}\partial_{\nu_2}\pi^{I_2}\cdots\partial_{\mu_n}\partial_{\nu_n}\pi^{I_{n}}\right) \ .
\ee

It should be remembered that all these theories containing Galilean-invariant operators are not renormalizable, i.e. they are effective field theories with some cutoff $\Lambda$, above which some UV completion is required.   But as mentioned, there can still be regions in which they dominate, since the symmetries forbid any renormalizable terms, and other terms will have more derivatives.  As was mentioned in the introduction, the ${\cal L}_n$ terms above, both in the single and multi-field case, do not get renormalized upon loop corrections, so that their classical values can be trusted quantum-mechanically \cite{Hinterbichler:2010xn}.

%%%%%%%%%%%%%%%%%%%%%%%%%%%%%

\section{Multi-Galileons and Higher co-Dimension Branes}

The Galilean symmetry can be thought of as inherited from symmetries of a probe brane floating in a higher dimensional flat bulk, in a small field limit \cite{deRham:2010eu}.  

The co-dimension one construction admits a generalization, that we, and others, have quite recently developed.  Consider co-dimension greater than one, and let the bulk coordinates be $X^A$, ranging over $D$ dimensions, and let the brane coordinates be $x^\mu$, ranging over $d$ dimensions, so that the co-dimension is $N=D-d$. 
The relevant action will still be invariant under the Poincare transformations 
\be 
\label{poincaretransformations} 
\delta_PX^A=\omega^A_{\ B}X^B+\epsilon^A \ ,
\ee
and the world-volume reparameterization gauge symmetries 
\be 
\label{gaugetransformations} 
\delta_g X^A=\xi^\mu\partial_\mu X^A \ ,
\ee
where $\xi^\mu(x)$ is the gauge parameter.  We may use this gauge freedom to fix a unitary gauge
\be 
X^\mu(x)=x^\mu,\ \ \ X^I(x)\equiv\pi^I(x) \ ,
\ee
where the $I$ part of the index $A$ represents directions transverse to the brane.  
The Poincare transformations~(\ref{poincaretransformations}) do not preserve this gauge, 
but it can be restored by making a compensating gauge transformation, $\delta_gX^\mu=\xi^\nu\partial_\nu x^\mu=\xi^\mu$, with the choice $
\xi^\mu=-\omega^\mu_{\ \nu}x^\nu-\omega^\mu_{\ I}\pi^I-\epsilon^\mu \ .$
Thus the combined transformation $\delta_{P'}=\delta_P+\delta_g$ leaves the gauge fixing intact and is a symmetry of the gauge fixed action.  Its action on the remaining fields $\pi^I$ is
\be \label{multiinternalpoincare}
\delta_{P'}\pi^I=-\omega^\mu_{\ \nu}x^\nu\partial_\mu\pi^I-\epsilon^\mu\partial_\mu \pi^I+\omega^I_{\ \mu}x^\mu-\omega^\mu_{\ J}\pi^J\partial_\mu\pi^I+\epsilon^I+\omega^I_{\ J}\pi^J \ .
\ee
The first five terms are obvious generalizations of those of the single Galileon theory, while the last term is new to co-dimension greater 
than one, and corresponds to the unbroken $so(N)$ symmetry in the transverse directions.  In total, the Poincare group $p(1,D-1)$ is broken to $p(1,d-1)\times so(N)$.  

Taking the small $\pi^I$ limit, we find the extended non-relativistic internal Galilean invariance under which the $\pi^I$ transform:  
\be
\label{multiinternalGalilean} \delta_{P'}\pi^I=\omega^I_{\ \mu}x^\mu+\epsilon^I+\omega^I_{\ J}\pi^J \ .
\ee
This consists of a Galilean invariance acting on each of the $\pi^I$ as in (\ref{generaltermspre}), and, importantly as we shall see, an extra internal $so(N)$ rotation symmetry under which the $\pi$'s transform as a vector.  

To obtain the multi-field actions invariant under (\ref{multiinternalGalilean}), we must choose the tensor $S$ in (\ref{generaltermspre}) so that it is invariant under $so(N)$ rotations acting on all its indices.  Equivalently, we must contract up the $I,J,\ldots$ indices on the fields with each other using $\delta_{IJ}$, the only $so(N)$ invariant tensor (contracting with the epsilon tensor would give a vanishing action).
This simple fact immediately rules out all the Lagrangians with an odd number of $\pi$ fields, including the DGP cubic term.  For an even number of $\pi$ fields, a little thought concerning the allowed ways to contract the Lorentz indices reveals that the unique multi-field Galileon can be written as
\bea
{\cal L}_{n+1}=n\eta^{\mu_1\nu_1\mu_2\nu_2\cdots\mu_n\nu_n}\left( \partial_{\mu_1}\pi^{I_1}\right.&&\partial_{\nu_1}\pi_{I_1}\partial_{\mu_2}\partial_{\nu_2}\pi^{I_2} \partial_{\mu_3}\partial_{\nu_3}\pi_{I_2}\cdots \nonumber \\
&&\left.\partial_{\mu_{n-1}}\partial_{\nu_{n-1}}\pi^{I_{n-1}}\partial_{\mu_n}\partial_{\nu_n}\pi_{I_{n-1}}\right) \ .
\eea

In four dimensions, there are now therefore only two possible terms; the kinetic term and a fourth order interaction term.
\bea  
{\cal L}_2&=& \partial_\mu\pi^I\partial^\mu\pi_I, \\
 {\cal L}_4&=&  \partial_\mu\pi^I\partial_\nu \pi_I\left(\partial^\mu\partial_\rho\pi^J\partial^\nu\partial^\rho\pi_J-\partial^\mu\partial^\nu\pi^J\square\pi_J\right) \nonumber \\
&&\ \ \ \ \ \ \ +{1\over 2}  \partial_\mu\pi^I\partial^\mu \pi_I\left(\square\pi^J\square\pi_J-\partial_\nu\partial_\rho\pi^J\partial^\nu\partial^\rho\pi_J\right) \ . 
 \nn 
 \label{multi4thorder}
 \eea
 
This represents an intriguing four dimensional scalar field theory: there is a single possible interaction term, and thus a single free coupling constant  (as in, for example, Yang-Mills theory).  Of course there are other possible terms compatible with the symmetries, namely those which contain two derivatives on every field, and where the field indices are contracted.  However, the quartic term above is the only one with six derivatives and four fields.  All other Galilean-invariant terms have at least two derivatives per field.   Thus, as argued in the introduction, there can exist regimes in which the above quartic term is the only one which is important.  Furthermore, this term is not renormalized to any order in perturbation theory, so classical calculations in these interesting regimes are in fact exact.  

To fully specify the theory, it is necessary to couple the $\pi$ fields to matter. The simple linear coupling $\pi^I T$, where $T\equiv \eta_{\mu\nu}T^{\mu\nu}$ is the trace of the energy momentum tensor, does not respect the $so(N)$ symmetry of the multi-Galileon Lagrangian. There are, of course, many other
couplings that do respect this symmetry. The simplest of these is $\pi^I\pi_I T$, but this has its own drawback, namely that it does not respect the Galilean symmetry. To leading order in an expansion in $\pi^I$, a coupling that respects both the internal $so(N)$ symmetry and the Galilean symmetry is given by
\be
\partial_{\mu}\pi^I\partial_{\nu}\pi_I T^{\mu\nu}_{\rm flat} \ ,
\ee
where $ T^{\mu\nu}_{\rm flat}$ is the energy-momentum tensor computed using the flat $4$-dimensional metric $\eta_{\mu\nu}$. Indeed such a coupling will
naturally emerge via the brane construction from a minimal coupling ${\cal L}_{\rm matter}(g_{\mu\nu}, \psi)$ to brane matter $\psi$.  Spherical solutions and constraints in the presence of some of these couplings are examined in \cite{Andrews:2010km}.

\subsection{Higher Co-dimension Branes and Actions}

In~\cite{deRham:2010eu}, it was shown how, in co-dimension one, to construct Galilean and internally relativistic invariant scalar field actions from the probe-brane prescription.  It is interesting to see how that approach can be generalized to higher co-dimensions.  

To begin, let us review the co-dimension $1$ construction. To obtain an action invariant under the Galilean symmetry we need only construct an action for the embedding of a brane $X^A(x),$ which is invariant under reparametrizations and Poincare transformations.  The reparametrizations force the action to be a diffeomorphism scalar constructed out of the induced metric $g_{\mu\nu}\equiv {\partial X^A\over\partial x^\mu} {\partial X^B\over\partial x^\nu} G_{AB}(X)$, where $G_{AB}$ is the bulk metric as a function of the embedding variables $X^A$.  Poincare invariance then requires the bulk metric to be the flat Minkowski metric $G_{AB}(X)=\eta_{AB}$.  Fixing the gauge $X^\mu(x)=x^\mu$ then fixes the induced metric
\be
 g_{\mu\nu}=\eta_{\mu\nu}+\partial_\mu \pi\partial_\nu\pi \ .
\ee
Any action which is a diffeomorphism scalar, $F$, evaluated on this metric, will yield an action for $\pi$ having the single field version of the internal Poincare 
invariance~(\ref{multiinternalpoincare}).  The ingredients available to construct such an action are the metric $g_{\mu\nu}$, the covariant derivative $\nabla_\mu$ compatible with the induced metric, the Riemann curvature tensor $R^{\rho}_{\ \sigma\mu\nu}$ corresponding to this derivative, and the extrinsic curvature $K_{\mu\nu}$ of the embedding.  Thus, the most general action is
\be
S=\left. \int d^4x\ \sqrt{-g}F\left(g_{\mu\nu},\nabla_\mu,R^{\rho}_{\ \sigma\mu\nu},K_{\mu\nu}\right)\right|_{g_{\mu\nu}=\eta_{\mu\nu}+\partial_\mu \pi\partial_\nu\pi} \ .
\ee
For example, the DBI action arises from
\be  
\int d^4x\ \sqrt{-g}\rightarrow  \int d^4x\ \sqrt{1+(\partial\pi)^2} \ .
\ee

To recover a Galilean-invariant action, with the symmetry~(\ref{Galileoninvarianceold}), we have only to take the small $\pi$ limit.  For example, the DBI action 
above yields the kinetic term ${\cal L}_2$ in this limit.  The DGP cubic term comes from the action $\sim \sqrt{-g}g^{\mu\nu}K_{\mu\nu}$.  Note that this in this construction the brane is merely a probe brane and no de-coupling limit is taken, which is fundamentally different from what occurs in the de-coupling limit of DGP.

To generalize this prescription to higher co-dimension, we must now consider diffeomorphism scalars constructed from the induced metric 
\be 
g_{\mu\nu}=\eta_{\mu\nu}+\partial_\mu \pi^I\partial_\nu\pi_I \ .
\ee
A much more difficult question concerns the ingredients from which to construct the action; i.e. the geometric quantities associated with a higher co-dimension brane.  This is a technical question, but here, let us point out that the main difference from the co-dimension one
case is that the extrinsic curvature now carries an extra index, $K^i_{\mu\nu}$.  The $i$ index runs over the number of co-dimensions, and is associated with an orthonormal basis in the normal bundle to the brane.  In addition, the covariant derivative $\nabla_\mu$ 
has a connection, $\beta^i_{\mu j}$, called the twist connection, that acts on the $i$ index.  For example, the covariant derivative of the extrinsic curvature reads 
\be
\nabla_\rho K^i_{\mu\nu}=\partial_\rho K^i_{\mu\nu}-\Gamma^\sigma_{\rho\mu}K^i_{\sigma\nu}-\Gamma^\sigma_{\rho\nu}K^i_{\mu \sigma}+\beta^i_{\rho j}K^j_{\mu\nu} \ .
\ee
The connection $\beta^i_{\mu j}$ is anti-symmetric in its $i,j$ indices, and so is a new feature appearing in co-dimensions $\geq2$; it vanishes in co-dimension one.  It has an associated curvature, $R^i_{\ j\mu\nu}$.  Therefore, an action of the form
\be
\label{generalmultiaction} 
S=\left. \int d^4x\ \sqrt{-g}F\left(g_{\mu\nu},\nabla_\mu,R^{i}_{\ j\mu\nu},R^{\rho}_{\ \sigma\mu\nu},K^i_{\mu\nu}\right)\right|_{g_{\mu\nu}=\eta_{\mu\nu}+\partial_\mu \pi^I\partial_\nu\pi_I} \ ,
\ee
will have the required relativistic symmetry~(\ref{multiinternalpoincare}), and its small field limit will have the Galilean invariance (\ref{multiinternalGalilean}).

\subsubsection{Brane quantities}

To evaluate the action~(\ref{generalmultiaction}), it is necessary to know how to express the various geometric quantities in terms of the $\pi^I$.  

The tangent vectors to the brane are 
\be 
e^A_{\ \mu}={\partial X^A\over \partial x^\mu}=\left\{\begin{array}{cr} \delta^\nu_\mu & A=\nu \ , \\ \partial_\mu\pi^I & A=I \ ,\end{array}\right.
\ee
and the induced metric is
\be 
g_{\mu\nu}=e^A_{\ \mu}e^B_{\ \nu}\eta_{AB}=\eta_{\mu\nu}+\partial_\mu\pi^I\partial_\nu\pi_I \ ,
\ee
where the $I$ index is raised and lowered with $\delta_{IJ}$.  The inverse metric can then be written as a power series,
\be 
g^{\mu\nu}=\eta^{\mu\nu}-\partial^\mu\pi^I\partial^\nu\pi_I+\mathcal{O}(\pi^4) \ .
\ee
%
%To find the (orthonormal) normal vectors $n^A_{\ i}$ (the index $i$ takes the same values as $I$, but it is the orthonormal frame index, whereas $I$ is the transverse coordinate index), we solve the defining equations
%\be 
%\label{normalvectorequations} 
%e^A_{\ \mu}n^B_{\ i}\eta_{AB}=0,\ \ \ n^A_{\ i}n^B_{\ j}\eta_{AB}=\delta_{ij} \ .
%\ee
%The first equation tells us that
%\be 
%n_{Ai}=\left\{\begin{array}{cr}-n_{Ii}\partial_\mu\pi^I & A=\mu \ , \\ n_{Ii} & A=I \ ,\end{array}\right. 
%\ee
%where $n_{Ii}$ are the as yet undetermined $A=I$ components of $n_{Ai}$.  The second equation of (\ref{normalvectorequations}) then gives
%\be 
%\label{transversevielbein} 
%\delta_{ij}=n^I_{\ i} n^J_{\ j}\left(\partial_\mu \pi_I\partial^\mu\pi_J+\delta_{IJ}\right) \ .
%\ee
%Thus, t
The $n^I_{\ i}$ must be chosen to be vielbeins of the transverse ``metric'' $g_{IJ}\equiv\partial_\mu \pi_I\partial^\mu\pi_J+\delta_{IJ}$.  The ambiguity in this choice due to local $O(N)$ transformations reflects the freedom to change orthonormal basis in the normal space of the brane.   The vielbeins summed over their Lorentz indices $i,j$ give the inverse of the metric to $g_{IJ}$, which expanded in powers of $\pi$ gives
\be 
n^I_{\ i}n^{J}_{\ j}\delta^{ij}=\delta^{IJ}-\partial_\mu\pi^I\partial^\mu\pi^J+\mathcal{O}(\pi^4) \ .
\ee
The metric determinant can be expanded as
\be 
\sqrt{-g}=1+\half \partial_\mu\pi^I\partial^\mu\pi_I+\mathcal{O}(\pi^4) \ ,
\ee
and the extrinsic curvature and the twist connection are
\be
K_{i\mu\nu}=-n_{Ii}\partial_\mu\partial_\nu \pi^I \ ,
\ee

\be
\beta_{\mu ij}= \left(\delta^{IJ}+\partial_\nu \pi^I \partial^\nu \pi^J\right)n_{Ii}\partial_\mu n_{Jj}+n_{Ii}n_{Jj}\partial^\nu \pi^I\partial_\mu\partial_\nu\pi^J \ ,
\ee
respectively. The action~(\ref{generalmultiaction}) is an $so(N)$ scalar, and so will not depend on how the $n^I_{\ i}$ are chosen.

\subsubsection{Lovelock terms and the probe brane prescription}

A general choice for the action~(\ref{generalmultiaction}) will not lead to scalar field equations that are second order.  Those actions that do lead to second order equations are precisely those that come from Lovelock invariants~\cite{Lovelock:1971yv} and their boundary terms.  The Lovelock invariants  are combinations of powers of the Riemann tensor which are dimensional continuations of characteristic classes.  The problem of finding extensions of the $\pi$ Lagrangian which possess second-order equations of motion is therefore
equivalent to the problem of finding extensions of higher-dimensional 
Einstein gravity which have second-order equations of motion.  In the presence of lower-dimensional hypersurfaces or branes, 
Einstein gravity in the bulk must be supplemented by the Gibbons-Hawking-York boundary term~\cite{Gibbons:1976ue,York:1972sj}.
This additional surface term is required in order to ensure that the variational problem of the combined brane/bulk system is well posed \cite{Dyer:2008hb}.
Similar boundary terms (Myers terms)~\cite{Myers:1987yn,Miskovic:2007mg} exist for the Lovelock invariants, and also lead to second order equations for $\pi$. (Note that the Galileon terms themselves need boundary terms in the presence of boundaries in 4d \cite{Dyer:2009yg}.)  

The prescription of~\cite{deRham:2010eu} is as follows:  the $d$-dimensional single field Galileon terms with an even number $N$ of $\pi$'s are obtained from the $(N-2)$-th Lovelock term on the brane, constructed from the brane metric. The terms with an odd number $N$ of $\pi$'s are obtained from the boundary term of the $(N-1)$-th $d+1$ dimensional bulk Lovelock term.  For instance, in $d=4$, the kinetic term with two $\pi$'s is obtained from $\sqrt{-g}$ on the brane; the cubic $\pi$ term is obtained from the Gibbons-Hawking-York term $\sqrt{-g}K$; the quartic term is obtained from $\sqrt{-g}R$; and the quintic term arises from the boundary term of the bulk Gauss-Bonnet invariant.  There are no further non-trivial Lovelock terms for $d=4$, in either the brane or the bulk, corresponding to the fact that there are no further non-trivial Galileon terms.

To extend this to higher co-dimension, we needed the corresponding higher-co-dimension boundary terms induced by the bulk Lovelock invariants.   These were studied by Charmousis and Zegers~\cite{Charmousis:2005ey}, who found that, despite
the freedom to specify a fairly general bulk gravitational theory and
number of extra dimensions, the resulting four-dimensional 
terms are surprisingly constrained, corresponding to the fact that the multi-Galileon action is essentially unique.  

For a brane of dimension $d=4$, the prescription depends on whether the co-dimension $N$ is odd or even. If $N\neq 3$ is odd,  one obtains the dimensional continuation of the Gibbons-Hawking-York and Myers terms, with the extrinsic curvature replaced by a distinguished normal component of $K^i_{\mu\nu}$.  In the special case $N=3$, there are additional terms involving the extrinsic curvature and the boundary term is not the dimensional continuation of the Myers term. If $N\neq 2$ is even, the boundary term includes only a brane cosmological constant and an induced Einstein-Hilbert term.  In the special case $N=2$, the boundary terms include only 
a brane cosmological constant, and the following term
\be\label{e:Neven2}
{\cal L}_{N=2} =
\sqrt{-g} \left(R[g] - (K^i)^2 + K_{\mu\nu}^i K^{\mu\nu}_{i}
 \right) \ .
\ee

\subsubsection{Recovering the multi-field Galileon}

For simplicity, we shall restrict ourselves to the even co-dimension case. The unique brane action in four dimensions for even co-dimension $\geq 4$ is then
\be 
S=\int d^4 x\ \sqrt{-g}\left(-a_2+a_4 R\right) \ .
\ee
The Galileon action is obtained by substituting $g_{\mu\nu}=\eta_{\mu\nu}+\partial_\mu\pi^I\partial_\nu\pi_I$, and expanding each term to lowest non-trivial order in $\pi$.  The cosmological constant term yields an $\mathcal{O}(\pi^2)$ piece, and the Einstein-Hilbert term yields an $\mathcal{O}(\pi^4)$ piece.  Up to total derivatives, we then have 
\be 
\label{fourthorderaction} 
S=\int d^4 x\ \left[-a_2 \ \half\partial_\mu\pi^I\partial^\mu\pi_I  +a_4\ \partial_\mu\pi^I\partial_\nu\pi^J\left(\partial_\lambda\partial^\mu\pi_J\partial^\lambda\partial^\nu\pi_I-\partial^\mu\partial^\nu\pi_I\square\pi_J \right)\right] \ .
\ee
By adding a total derivative, it is straightforward to see that the $a_4$ term is proportional to the fourth order term~(\ref{multi4thorder}), and so we recover the four dimensional multi-field Galileon model
\be 
S=\int d^4 x\ \left[-{1\over 2} a_2  {\cal L}_2+{1\over 2} a_4  {\cal L}_4\right] \ .
\ee

The equations of motion are
\bea 
{\delta S\over \delta \pi^I}=  a_2\square \pi_I &+& a_4\left[\square\pi_I \left(\partial_\mu\partial_\nu\pi_J\partial^\mu\partial^\nu\pi^J- \square\pi^J\square\pi_J\right)\right. \nonumber \\
&&\ \ \ \ \ + \left. 2\partial_\mu\partial_\nu\pi_I \left(  \partial^\mu\partial^\nu\pi_J\square\pi^J -\partial^\mu\partial_\lambda\pi_J\partial^\nu\partial^\lambda\pi^J\right)\right] \ . \nn \\
\label{multi4eom}
\eea

For co-dimension two, there is the additional $K^2$ part to the boundary term.  This cancels the contribution from the Ricci scalar, and thus yields nothing new.  Therefore,~(\ref{fourthorderaction}) is the unique multi-Galileon term in four dimensions and any even co-dimension.  Some properties of the maximally symmetric solutions of multi-galileons, including an interesting kinetic Goldstone-like theorem, are studied in \cite{Hinterbichler:2010xn,Zhou:2010di}.  Keeping all orders in $\pi$ leads to a relativistically invariant action, a multi-field generalization of DBI with second order equations.

%%%%%%%%%%%%%%%%%%%%%%%%%%%%%

\section{Symmetries for Galileons on Curved Spaces}

While the Galileons have fascinating properties, if they are to be useful for cosmology, or form part of a more complete and dynamical picture, it is natural to consider extending them to a covariant form.  Performing this in a straightforward manner using minimal coupling ruins the second order property of the equations of motion.  Non-minimal curvature terms can be added to restore the second order property, but at the expense of the Galilean symmetry \cite{Deffayet:2009mn,Deffayet:2009wt,Deffayet:2011gz}.  However, in recent
papers~\cite{Goon:2011qf,Goon:2011uw} we have demonstrated how to construct such models while preserving the symmetry\footnote{This construction has connections to some other recently proposed ones~\cite{Deffayet:2011gz,Burrage:2011bt}}.

The general context is the theory of a dynamical 3-brane moving in a fixed (4+1)-dimensional background, but we now allow the bulk metric to be general, rather than flat Minkowski.  
The dynamical variables are the brane embedding $X^A(x)$, five functions of the world-volume coordinates $x^\mu$.  The bulk has the arbitrary but fixed background metric $G_{AB}(X)$, from which we may construct the induced metric $\bar g_{\mu\nu}(x)$ and the extrinsic curvature $K_{\mu\nu}(x)$
\bea 
\bar g_{\mu\nu}&=&e^A_{\ \mu}e^B_{\ \nu} G_{AB}(X) \ , \\ 
K_{\mu\nu}&=&e^A_{\ \mu}e^B_{\ \nu}\nabla_A n_B \ ,
\eea
where $e^A_{\ \mu}= {\partial X^A\over\partial x^\mu}$ are the tangent vectors to the brane, and $n^A$ is the unit normalized normal vector.

Generalizing the brane construction of \cite{deRham:2010eu}, the world-volume action must be gauge invariant under reparametrizations of the brane, which is guaranteed if the action is written as a diffeomorphism scalar, $F$, of $\bar g_{\mu\nu}$, $K_{\mu\nu}$, the covariant derivative $\bar\nabla_\mu$ and the curvature $\bar R^\alpha_{\ \beta\mu\nu}$ constructed from $\bar g_{\mu\nu}$.
\be
\label{generalaction} 
S= \int d^4x\ \sqrt{-\bar g}F\left(\bar g_{\mu\nu},\bar\nabla_\mu,\bar R^{\alpha}_{\ \beta\mu\nu},K_{\mu\nu}\right) \ .
\ee
This action possesses global symmetries only if the bulk metric possesses Killing symmetries. We fix all the gauge symmetry by choosing the gauge
\be
\label{physgauge} 
X^\mu(x)=x^\mu, \ \ \ X^5(x)\equiv \pi(x) \ ,
\ee
where we have foliated the bulk by time-like slices given by the surfaces $X^5(x)= {\rm constant}$.  The remaining coordinates $X^\mu$ can then be chosen arbitrarily and parametrize the leaves of the foliation. In this gauge, the coordinate $\pi(x)$ measures the transverse position of the brane relative to the foliation, and the resulting action solely describes $\pi$,
\be
\label{gaugefixedaction} 
S= \int d^4x\ \left. \sqrt{-\bar g}F\left(\bar g_{\mu\nu},\bar\nabla_\mu,\bar R^{\alpha}_{\ \beta\mu\nu},K_{\mu\nu}\right)\right|_{X^\mu=x^\mu,\ X^5=\pi} \ .
\ee

Since gauge fixing cannot alter global symmetries, any global symmetries of~(\ref{generalaction}) become global symmetries of~(\ref{gaugefixedaction}).   However, the form of the global symmetries depends on the gauge because the gauge choice is not generally preserved by the global symmetry.  Given a transformation generated by a Killing vector, $K^A$, we restore our preferred gauge (\ref{physgauge}) by making a compensating gauge transformation $\delta_{g,{\rm comp}}x^\mu=-K^\mu$.  The two symmetries then combine to shift $\pi$ by
\be
\label{gaugefixsym} 
(\delta_K+\delta_{g,{\rm comp}})\pi=-K^\mu(x,\pi)\partial_\mu\pi+K^5(x,\pi) \ ,
\ee
which is a symmetry of the gauge fixed action~(\ref{gaugefixedaction}).

%%%%%%%%

It is convenient at this stage to make two simplifying assumptions. We specialize to the case in which the foliation is Gaussian normal with respect to the metric $G_{AB}$, and we demand that the extrinsic curvature on each of the slices be proportional to the induced metric. Under these assumptions the metric takes the form
\be 
\label{metricform} 
G_{AB}dX^AdX^B=d\rho^2+f(\rho)^2g_{\mu\nu}(x)dx^\mu dx^\nu \ ,
\ee
where $X^5=\rho$ denotes the transverse coordinate, and $g_{\mu\nu}(x)$ is an arbitrary brane metric.  This special case includes all examples in which a maximally symmetric ambient space is foliated by maximally symmetric slices.  Killing symmetries which preserve the foliation will be linearly realized, whereas those that don't are realized nonlinearly.  Thus, the algebra of all Killing vectors is spontaneously broken to the subalgebra of Killing vectors preserving the foliation.

In the gauge (\ref{physgauge}), the induced metric is $\bar g_{\mu\nu}=f(\pi)^2g_{\mu\nu}+\nabla_\mu\pi\nabla_\nu\pi$.
Defining the quantity $\gamma=1/ \sqrt{1+{1\over f^2}(\nabla\pi)^2}$,
the extrinsic curvature is then
\be 
K_{\mu\nu}=\gamma\left(-\nabla_\mu\nabla_\nu\pi+f f'g_{\mu\nu}+2{f'\over f}\nabla_\mu\pi\nabla_\nu\pi\right) \ .
\ee

%A general choice for the action~(\ref{gaugefixedaction}) will lead to scalar field equations for $\pi$ which are higher than second order in derivatives and may therefore propagate extra ghost degrees of freedom.  
%However, as pointed out in~\cite{deRham:2010eu}, there are a finite number of actions that lead to second order equations.
%The possible
%extensions of Einstein gravity which remain second order are given by Lovelock terms and their boundary terms.  
%These terms are specific combinations of powers of the Riemann tensor which are topological (i.e. total derivatives) in some specific home dimension, but in lower dimensions have the property that equations of motions derived from them are second order.

%The prescription of~\cite{deRham:2010eu} is then as follows: 

On the 4-dimensional brane, we can add four Lovelock and boundary terms, plus a tadpole term,
%we may add the first two Lovelock terms, namely the cosmological constant term $\sim \sqrt{-\bar g}$ and the Einstein-Hilbert term $\sim \sqrt{-\bar g}R[\bar g]$.  (The higher Lovelock terms will be total derivatives in 4-dimensions.)  We may also add the boundary term corresponding to a bulk Einstein-Hilbert term, $\sqrt{-\bar g}K$, and the boundary term ${\cal L}_{\rm GB}$ corresponding to the Gauss-Bonnet Lovelock invariant $R^2 - 4 R_{\mu\nu} R^{\mu\nu}+ R_{\mu\nu\alpha\beta} R^{\mu\nu\alpha\beta}$ in the bulk.  The zero order cosmological constant Lovelock term in the bulk has no boundary term, although we may construct a tadpole-like term from it, and the higher order bulk Lovelock terms vanish identically.  Therefore, in total, for a 3-brane there are five terms that lead to second order equations for $\pi$,
\bea   {\cal L}_1&=&\sqrt{-g}\int^\pi d\pi' f(\pi')^4,\nn\\
{\cal L}_2&=&- \sqrt{-\bar g} \ ,\nn\\
{\cal L}_3&=& \sqrt{-\bar g}K \ ,\nn\\
{\cal L}_4&=& -\sqrt{-\bar g}\bar R \ ,\nn\\
{\cal L}_5&=&{3\over 2}\sqrt{-\bar g} {\cal K}_{\rm GB} \ ,
\label{ghostfreegenterms} \eea
where the explicit form of the Gauss-Bonnet boundary term is
${\cal K}_{\rm GB}=-{1\over3}K^3+K_{\mu\nu}^2K-{2\over 3}K_{\mu\nu}^3-2\left(\bar R_{\mu\nu}-\half \bar R \bar g_{\mu\nu}\right)K^{\mu\nu}$. 

$\mathcal L_1$ is the zero derivative tadpole term which is the proper volume between any $\rho=$ constant surface and the brane position, $\pi(x)$ \cite{Goon:2011qf}.  While different in origin from the other terms, it too has the symmetry~(\ref{gaugefixsym}).  Each of these terms may appear in a general Lagrangian with an arbitrary coefficient. 

Evaluating these expressions for the metric~(\ref{metricform}) involves a lengthy calculation which ultimately yields
\bea   
{\cal L}_1&=&\sqrt{-g}\int^\pi d\pi' f(\pi')^4,\nn\\
{\cal L}_2&=&-\sqrt{-g}f^4\sqrt{1+{1\over f^2}(\partial\pi)^2},\nn\\
{\cal L}_3&=&\sqrt{-g}\left[f^3f'(5-\gamma^2)-f^2[\Pi]+\gamma^2[\pi^3]\right],\nn \\
{\cal L}_4&=& -\sqrt{-g}\Big\{{1\over\gamma}f^2R-2{\gamma}R_{\mu\nu}\nabla^\mu\pi\nabla^\nu\pi \nn\\
&&\ \ +\gamma\left[[\Pi]^2-[\Pi^2]+2{\gamma^2\over f^2}\left(-[\Pi][\pi^3]+[\pi^4]\right)\right]\nn\\
&&\ \ \ \!+6{f^3f''\over \gamma}\left(-1+\gamma^2\right) \!	 \nn \\
&&\ \ \ \ +2\gamma ff'\left[-4[\Pi]+{\gamma^2\over f^2}\left(f^2[\Pi]+4[\pi^3]\right)\right]\nn\\
&&\ \ \ \ \ -6{f^2f'^2\over \gamma}\left(1-2\gamma^2+\gamma^4\right) \Big\}.
\eea
The expression for ${\cal L}_5$ is lengthy, and can be found in~\cite{Goon:2011qf}.

In these expressions, all curvatures and covariant derivatives are those of the background metric $g_{\mu\nu}$.  The notation is as earlier, but with covariant derivatives replacing partial ones. The equations of motion contain no more than two derivatives on each field, ensuring that no extra degrees of freedom propagate around any background.

There are interesting cases in which the 5d background metric has 15 global symmetries, the maximal number-- the bulk can be either $5$d anti-de Sitter space $AdS_5$ with isometry algebra $so(4,2)$, 5d de-Sitter space $dS_5$ with isometry algebra $so(5,1)$, or flat 5d Minkowski space $M_5$ with isometry algebra the five dimensional Poincare algebra $p(4,1)$.  In addition, there are cases where the brane metric $g_{\mu\nu}$, and hence the extrinsic curvature, are maximally symmetric, so that the unbroken subalgebra has the maximal number of generators, 10.  This means that the leaves of the foliation are either $4$d anti-de Sitter space $AdS_4$ with isometry algebra $so(3,2)$, 4d de-Sitter space $dS_4$ with isometry algebra $so(4,1)$, or flat 4d Minkowski space $M_4$ with isometry algebra the four dimensional Poincare algebra $p(3,1)$.  In fact, there are only 6 such possible foliations of $5$d maximally symmetric spaces by $4$d maximally symmetric time-like slices, such that the metric takes the form~(\ref{metricform}).  Flat $M_5$ can be foliated by flat $M_4$ slices or by $dS_4$ slices; $dS_5$ can be foliated by flat $M_4$ slices, $dS_4$ slices, or $AdS_4$ slices; and $AdS_5$ can only be foliated by $AdS_4$ slices.  Each of these 6 foliations, through the construction leading to~(\ref{gaugefixedaction}), will generate a class of theories living on an $AdS_4$, $M_4$ or $dS_4$ background and having 15 global symmetries broken to the 10 isometries of the brane, the same numbers as those of the original galileon.  

To develop the analogues of the original Galileon theory and simplify the actions, we expand the Lagrangians in powers of $\lambda$ around some constant background, $\pi\rightarrow\pi_0+\lambda\pi$.  One can find appropriate linear combinations of the Lagrangians, $\bar{\mathcal L}_n=c_1\mathcal L_1+\ldots +c_n\mathcal L_n$, for which all terms $\mathcal{O}\left (\lambda^{n-1}\right )$ or lower are total derivatives.  When this prescription is carried out for the remaining four maximally symmetric cases in which the 4d background is curved, new classes of theories are produced.  After canonical normalization, $\hat\pi={1\over L^2}\pi$ where $L$ is the $dS_4$ or $AdS_4$ radius, the Lagrangians become
\begin{eqnarray} 
\hat{\cal L}_1&=&\sqrt{-g}\hat\pi \ , \nn\\
\hat{\cal L}_2&=&-\half\sqrt{-g} \left((\partial\hat\pi)^2-{R\over 3}\hat \pi^2\right) \ ,\nn \\
\hat{\cal L}_3&=& \sqrt{-g}\left(-{1\over 2}(\partial\hat\pi)^2[\hat\Pi]-{R\over 4} (\partial\hat\pi)^2\hat\pi+{R^2\over 36}\hat\pi^3\right) \ ,\nn\\
\hat{\cal L}_4&=&\sqrt{-g}\Big[-\half(\partial\hat\pi)^2\Big([\hat\Pi]^2-[\hat\Pi^2]+{R\over 24}(\partial\hat\pi)^2\nn\\
&&\ \ \ \ \ \ \ \ \  \ \ \ +{R\over 2}\hat\pi[\hat\Pi]+{R^2\over 8}\hat\pi^2\Big)+{R^3\over 288}\hat\pi^4\Big] \ ,
\label{singlesetGalileons} 
\eea 
with ${\cal L}_5$ again found in~\cite{Goon:2011qf}.

Here $R=\pm{12\over L^2}$ is the Ricci curvature of the $dS_4$ or $AdS_4$ background. These simpler Lagrangians are Galileons that live on curved space yet retain the same number of symmetries as the full theory, whose form comes from expanding~(\ref{gaugefixsym}) in appropriate powers of $\lambda$.  In the case of a $dS_4$ background in conformal inflationary coordinates $(u,y^i)$, the non-linear symmetries are
\be \label{dSGalileontrans}
\delta_{+}\hat\pi={1\over u}\left(u^2-y^2\right) , \ \ \
\delta_{-} \hat\pi=-{1\over u},\ \ \ 
\delta_{i} \hat\pi = {y_i\over u} \ .
\ee

A striking feature of these fully covariant models which is not present in the flat space Galileon theories is the presence of potentials whose couplings are determined by the symmetries~(\ref{gaugefixsym}).  In particular, the scalar field acquires a mass of order the $dS_4$ or $AdS_4$ radius, with a value protected by the symmetries (\ref{dSGalileontrans}), so the small masses should be protected against renormalization.

%%%%%%%%%%%%%%%%%%%%%%%%%%%%%

\section{Summary}
The realization that branes are important fundamental objects within string theory has motivated entirely new approaches to the construction of phenomenological models relevant to both particle physics and cosmology. The DGP model is an important example
of this, and, in recent years, it has been realized that it allows the development of four-dimensional effective field theories with novel properties. These abstractions of
the original idea have proven fascinating in their own rights, and we have been hard at work over the past year attempting to understand how to construct the most general versions of them.

In this review we have described the original Galileon idea, and then discussed our work to extend it in two specific directions -- to multiple fields, and to a fully covariant and symmetric version. In both cases we have described the most general four-dimensional effective theory, and how it can arise though the embedding of a suitable brane in a suitable
ambient space. The resulting symmetries are then inherited from combinations of five-dimensional Poincar\'e invariance and brane reparametrization invariance. In the case of multiple Galileons, we have shown that, in addition to $N$ copies of the single field Galilean symmetry, an internal $so(N)$ symmetry emerges, reflecting rotational invariance of the brane bending modes in the isotropic flat extra dimensional manifold.  In the case of Galileons on a curved background, the story is much richer, with
maximally symmetric choices for the bulk and brane metrics yielding complicated four dimensional symmetry groups, describing potentials with non-renormalized coefficients.

Much of this work is very recent, and may point the way to novel applications in inflation, cosmic acceleration, and perhaps in particle physics model building. An interesting open question is whether these fascinating four dimensional effective field theories, constructed through the dynamics of branes in higher dimensional spaces, might find a natural embedding within string theory, from whence many of the original motivations arose.
%%%%%%%%%%%%%%%%%%%%%%%%%%%%%

\end{document}